\documentstyle[preprint,aps,epsf,floats,rotate]{revtex}

\begin{document}

\preprint{
\vbox{\hbox{JHU--TIPAC--96022}
      \hbox{MADPH-97-994}
      \hbox{hep-ph/9707365} } }

\title{Leptoproduction of $J/\psi$}
\author{Sean~Fleming~\footnote{address after Sept. 1
1997: University of Toronto, 60 St.~George St., Toronto, Ontario CA, 
M5S1A7} }
\address{Department of Physics, University of Wisconsin, Madison \\
Madison, Wisconsin 53706 U.S.A. \\
{\tt fleming@pheno.physics.wisc.edu} }
\author{Thomas~Mehen~\footnote{address after Sept. 1 1997:
California Institute of Technology, Pasadena, CA 91125} }
\address{Department of Physics and Astronomy, The Johns Hopkins
University \\
3400 North Charles Street, Baltimore, Maryland 21218 U.S.A. \\
{\tt mehen@dirac.pha.jhu.edu} }

\date{July 1997}

\maketitle
\begin{abstract}

We study leptoproduction of $J/\psi$ at large  $Q^2$ 
within the nonrelativistic QCD (NRQCD) factorization formalism.
The cross section is dominated by color-octet terms that are of order
$\alpha_s$. The color-singlet term, which is of order $\alpha^2_s$,
is shown to be a small contribution to the total cross section. We
also calculate the tree diagrams for color-octet
production at order $\alpha^2_s$ in a region of phase space where
there is no leading color-octet contribution. We find that in this
regime the color-singlet contribution dominates.  We argue that non-perturbative corrections arising from diffractive leptoproduction, higher twist effects, and higher order terms in the NRQCD velocity expansion
should be suppressed as $Q^2$ is increased.  Therefore, the color-octet matrix elements $\langle {\cal O}_8^{\psi}(^1S_0)\rangle$ and $\langle  
{\cal O}_8^{\psi}(^3P_0)\rangle$ can be reliably extracted from this  
process. Finally, we point out that an experimental measurement of the  
polarization of leptoproduced $J/\psi$ will provide an excellent  
test of the NRQCD factorization formalism.
\end{abstract}

\pagebreak

\section{Introduction}

Quarkonium production has been the focus of much experimental and
theoretical attention in recent years. This surge in interest is
largely due to the observation of gross discrepancies between
experimental measurements of $J/\psi$ and $\psi^{\prime}$ production
at the Collider Detector Facility (CDF) at the Fermilab
Tevatron~\cite{CDF} and
calculations based on the Color-Singlet Model~\cite{Schuler}.
The dramatic failure of the traditional technique for calculating
quarkonia production and decay rates has led to a new paradigm for
understanding quarkonia: the nonrelativistic QCD (NRQCD) factorization
formalism of Bodwin, Braaten, and Lepage~\cite{BBL}.

A central result of the NRQCD factorization formalism is that
inclusive quarkonium production cross sections have the form of a sum 
of products of short-distance coefficients and NRQCD matrix elements.  
The short-distance coefficients are associated with the production of 
a heavy quark-antiquark pair in specific color and angular-momentum 
states.  They can be calculated using ordinary perturbative
techniques.  The NRQCD matrix elements parameterize the
effect of long-distance physics such as the hadronization of the
quark-antiquark pair.  These can be determined phenomenologically.

The power of the NRQCD formalism stems {}from the fact that
factorization formulas for observables are expansions in
the small parameter $v$, where $v$ is the average relative velocity
of the heavy quark and anti-quark in the quarkonium bound
state. For charmonium $v^2\sim 0.3$, and for bottomonium $v^2\sim 0.1$.  
NRQCD $v$-scaling rules~\cite{lmnmh} allow us to estimate the relative 
sizes of various NRQCD matrix elements.  This information, along with 
the dependence of the short-distance coefficients on $\alpha_s$ and $\alpha$,
permits us to decide which terms must be retained in expressions for
observables to reach a given level of accuracy.  At low
orders, factorization formulas involve only a few matrix
elements, so several observables can be related by a small number of
parameters.

Prior to the innovations presented in Ref.~\cite{BBL}, most $J/\psi$
production calculations took into account only the hadronization of
$c\bar{c}$ pairs initially produced in a color-singlet ${}^3S_1$
state, as parameterized by the NRQCD matrix element
$\langle{\cal O}^\psi_1({}^3S_1)\rangle$.
An important aspect of the NRQCD formalism is
that, in addition to the color-singlet contribution, it allows for the
possibility that a $c\overline{c}$ pair produced in a color-octet
state can evolve nonperturbatively into a $J/\psi$ or
$\psi^{\prime}$. This color-octet mechanism is central to the current
theoretical understanding of charmonium production at the Tevatron.

For a majority of phenomenological applications of the NRQCD
factorization formalism, the most important color-octet matrix
elements are
$\langle{\cal O}^\psi_8({}^3S_1) \rangle$,
$\langle{\cal O}^\psi_8({}^1S_0) \rangle$, and
$\langle{\cal O}^\psi_8({}^3P_J) \rangle$.
They describe the non-perturbative evolution of a color-octet
$c\overline{c}$ pair in either a ${}^3S_1$, ${}^1S_0$, or ${}^3P_J$
angular momentum state into a $J/\psi$.
Using heavy quark spin symmetry
relations \cite{BBL}, it is possible to express all three P-wave
matrix elements in terms of one:
$\langle {\cal O}^\psi_8(^3P_J)\rangle =
(2J+1) \langle {\cal O}^\psi_8(^3P_0)\rangle + O(v^2)$. Thus, at
leading order in $v$, there are three color-octet matrix elements.
At this time they cannot be
computed from first principles, and must therefore be extracted from
experimental data.

The most precise determination of the color-octet matrix element
$\langle{\cal O}^\psi_8({}^3S_1) \rangle$ comes from a fit of a
leading order theoretical calculation~\cite{CL,BK} to CDF data on
$J/\psi$  production
at high transverse momentum ($P_{\perp}$). A
linear combination of the remaining two color-octet matrix elements 
$\langle{\cal O}^\psi_8({}^1S_0) \rangle$ and
$\langle{\cal O}^\psi_8({}^3P_0) \rangle$ can also be determined by 
fitting the leading order calculation to CDF data on $J/\psi$
produced at low to moderate
$P_{\perp}$. The fit is extremely
sensitive to theoretical uncertainties, so that the value determined 
for the linear combination of $\langle{\cal O}^\psi_8({}^1S_0) \rangle$ 
and $\langle{\cal O}^\psi_8({}^3P_0) \rangle$ can only be regarded as an 
order of magnitude estimate. There are other $J/\psi$ production  
processes,
such as those measured at fixed target experiments in 
photonic~\cite{Photo},
and hadronic~\cite{Hadro} collisions, that could, in
principle, provide a means to determine
$\langle{\cal O}^\psi_8({}^1S_0) \rangle$ and
$\langle{\cal O}^\psi_8({}^3P_0) \rangle$. However, in each case
there are large theoretical uncertainties that only allow for an order 
of magnitude estimate of these color-octet matrix elements.

It is clear that testing the NRQCD factorization formalism requires 
a precise determination of each of the leading matrix elements.
Specifically, the challenge lies in an accurate measurement of
$\langle{\cal O}^\psi_8({}^1S_0) \rangle$ and
$\langle{\cal O}^\psi_8({}^3P_0) \rangle$.
In this paper we study leptoproduction of $J/\psi$, and show
that it is possible to determine these matrix elements 
from a measurement of the unpolarized $J/\psi$ production cross
section. We also show that once these matrix elements are extracted,
the polarization of the $J/\psi$ can be predicted without introducing
any new parameters. Thus, a measurement of the polarization of
leptoproduced $J/\psi$ will provide an excellent test of the NRQCD
factorization formalism. 

An explicit calculation need not be carried
out to understand why it should be possible to determine
$\langle{\cal O}^\psi_8({}^1S_0) \rangle$ and
$\langle{\cal O}^\psi_8({}^3P_0) \rangle$
from a measurement of the
$J/\psi$ leptoproduction cross section. The $O(\alpha_s)$
contribution to $J/\psi$ production comes from processes in which
the $c\bar{c}$ pair is produced at short-distances in either a  
color-octet
${}^{1}S_{0}$ or ${}^{3}P_{J}$ configuration. It is not until one
includes the $O(\alpha_s^2)$ contributions that it is
possible for $J/\psi$ to be produced through the production at
short distances of a $c\bar{c}$ pair in a color-singlet ${}^{3}S_{1}$
state. The color-octet matrix elements $\langle{\cal O}^\psi_8({}^1S_0) \rangle$ and $\langle{\cal O}^\psi_8({}^3P_0) \rangle$ are suppressed 
by $v^{3}$ and $v^4$, respectively, relative to the color-singlet matrix element. Since the perturbative coefficient for color-octet production is enhanced relative to the perturbative coefficient for color-singlet production by a factor of $\sim \pi/\alpha_s$, we expect the color-octet contribution to be roughly $v^3 \pi/\alpha_s \approx 2$ times the color-singlet contribution. In fact, upon completing the calculation, we find that color-octet contribution is about $4$ times the size of the leading order color-singlet contribution.  
 
Note that since $Q^2$ can be large, leptoproduction is a better process
from which to extract
$\langle {\cal O}^{\psi}_8(^1S_0)\rangle$ and
$\langle {\cal O}^{\psi}_8(^3P_0)\rangle$ than photoproduction and 
low-energy hadroproduction.  The latter two processes
lack any large scale other than the charm quark mass, and consequently, 
perturbative corrections to leading order
calculations are larger. In addition nonperturbative effects, such as higher twist corrections to the parton model and diffractive $J/\psi$ production, 
are less effectively suppressed in photoproduction and low-energy hadroproduction than in leptoproduction at large $Q^2$.

Finally, we point out that there are $J/\psi$ leptoproduction final
states which can only be described by $O(\alpha_s^2)$ tree-level
contributions ({\it i.e.} there is no $O(\alpha_s)$ color-octet
contribution). The final states are those in which a gluon jet is
well separated from the $J/\psi$. Though the dominant contribution to
these final states comes from the $O(\alpha_s^2)$ color-singlet term,
there are also $O(\alpha_s^2)$ color-octet contributions. We calculate
these color-octet terms to determine if there is any region of phase
space where they are enhanced relative to the color-singlet term. We
find that in the regime where the calculation is valid the
color-singlet term always dominates.

This paper is organized in such a way as to separate theory from
phenomenology.  Readers who are only interested in the
phenomenological implications of this work can skip the theoretical 
discussion without loss of coherence.

We discuss theoretical issues in sections II--IV.  First, we briefly
review the NRQCD factorization formalism.
Then we present the $O(\alpha_s)$ and $O(\alpha_s^2)$ tree level calculations of $J/\psi$ leptoproduction.  At $O(\alpha_s)$, we calculate both the polarized and unpolarized $J/\psi$ production cross sections.
Finally, we discuss possible corrections to our calculations 
from diffractive production and higher twist corrections to the  
parton model. We also address the possibility of a breakdown of the  
NRQCD velocity expansion near the boundaries of phase space. For the distributions computed in this paper, we argue that these errors  
are systematically reduced as $Q^2$ is made large.

The phenomenology of $J/\psi$ leptoproduction is presented in sections
V and VI. First, we discuss the
present status of the determination of the leading order NRQCD matrix 
elements. Then we present our
predictions, and discuss the
theoretical limitations on the accuracy to which
$\langle {\cal O}^{\psi}_8(^1S_0)\rangle$ and
$\langle {\cal O}^{\psi}_8(^3P_0)\rangle$ can be measured in
leptoproduction. We conclude that leptoproduction of unpolarized
$J/\psi$ is an excellent way to measure
$\langle {\cal O}^{\psi}_8(^1S_0)\rangle$ and
$\langle {\cal O}^{\psi}_8(^3P_0)\rangle$, and that leptoproduction of
polarized $J/\psi$ provides a powerful test of the NRQCD factorization
formalism.

\section{The NRQCD Factorization Formalism}

The NRQCD factorization formalism of Bodwin, Braaten, and Lepage  
\cite{BBL}
has emerged as a new paradigm for computing the
production and decay rates of heavy quarkonia. This formalism
provides a rigorous theoretical framework which systematically
incorporates relativistic corrections and ensures the infrared safety of 
perturbative calculations~\cite{IR}. In the NRQCD factorization formalism,
cross sections for the production of a quarkonium state $H$ are written 
as
\begin{equation}
\label{Factor}
\sigma(H) =
\sum_n {F_n \over m^{d_{n}-4}_{Q}} \langle 0|{\cal O}_n^H|0\rangle ,
\end{equation}
where $m_{Q}$ is the mass of the heavy quark $Q$.
The short-distance coefficients, $F_n$, are associated with the
production, at distances of order $1/m_{Q}$ or less, of a $Q\bar{Q}$ pair 
with quantum numbers indexed by $n$ (angular momentum ${}^{2S+1}L_{J}$ 
and color $1$ or $8$). They are computable in perturbation theory.
In Eq.~(\ref{Factor}), $\langle 0| {\cal O}_n^H|0 \rangle$ are  
vacuum matrix
elements of NRQCD operators:
\begin{equation}
\label{On}
\langle 0| {\cal O}_n^H|0 \rangle \equiv \sum_X \sum_{\lambda}
\langle 0|{\cal K}_n^{\dagger}|H(\lambda) + X\rangle
\langle H(\lambda) + X |{\cal K}_n|0 \rangle,
\end{equation}
where ${\cal K}_n$ is a bilinear in heavy quark fields
which creates a $Q \bar{Q}$ pair in an angular-momentum and color
configuration indexed by $n$. The bilinear combination
${\cal K}_n^{\dagger}{\cal K}_n$ has energy dimension $d_{n}$.
The production matrix elements describe the
evolution of the $Q \bar{Q}$ pair into a final state containing the 
quarkonium $H$ plus additional hadrons ($X$) which are soft in the
quarkonium rest frame.

Throughout the
remainder of the paper, we use a shorthand notation in which the vacuum 
matrix elements are written as
$\langle {\cal O}^H_{(1,8)}(^{2S+1}L_J) \rangle$.

The NRQCD
matrix elements obey simple scaling laws~\cite{lmnmh} with respect
to $v$, the relative
velocity of the $Q$ and $\bar{Q}$. Therefore, Eq.~(\ref{Factor})
is a double expansion in $v$ and $\alpha_s$. Since the NRQCD matrix elements
are sensitive only to large distance scales they are independent of 
the short-distance process in which the $Q$ and $\bar{Q}$ are
produced. Thus, it is possible to extract numerical values for the
NRQCD  matrix elements in one experiment and use them to predict  
production
cross sections in other processes.

\section{Leptoproduction Calculation}

\begin{figure}
\epsfxsize=10cm
\hfil\epsfbox{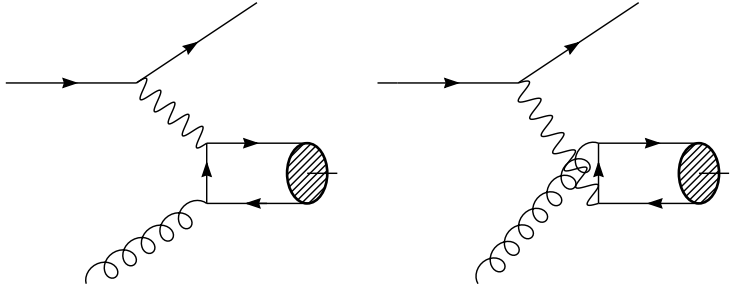}\hfill
\caption{Leading order diagrams for color-octet leptoproduction of
$J/\psi$}
\label{Leading}
\end{figure}

At leading order in $\alpha_{s}$, $J/\psi$ are produced 
through the hadronization of a
$c\bar{c}$ pair in either a ${}^1S_0$, or ${}^3P_J$ configuration. The 
Feynman diagrams are shown in Fig.~\ref{Leading}.

Theoretical predictions for the $J/\psi$ leptoproduction cross
section can be computed using the techniques of Ref. \cite{BC}. The 
short-distance coefficients appearing in Eq.~(\ref{Factor}) can be
computed by matching a perturbative calculation in full QCD with a
corresponding perturbative calculation in NRQCD. The production cross 
section of a $c \overline{c}$ pair with relative three-momentum
${\bf q}$ is computed in full QCD, and Taylor expanded in powers of
${\bf q}$.  In this Taylor expansion, the four-component Dirac spinors
are expressed in terms of nonrelativistic two-component heavy quark 
spinors. The NRQCD matrix elements on the right-hand side of
Eq.~(\ref{Factor}) are easily expressed in terms of the two-component 
heavy quark spinors and powers of ${\bf q}$. The $F_n$
appearing in Eq.~(\ref{Factor}) are then chosen so that the full QCD 
calculations and NRQCD calculations agree. Detailed examples of the 
calculational technique can be found in Ref.~\cite{BC}; here we will 
merely quote our result for the differential cross section.

The expression for the cross section determined from the diagrams in
Fig.~\ref{Leading} is
\begin{eqnarray}
\lefteqn{
\sigma(e+p\to e+\psi +X) = \int {dQ^2 \over Q^2}\int dy
\int dx\; f_{g/p}(x)\; \delta(xys-(2m_c)^{2}-Q^2) }
\nonumber \\
& & \;\;\;\;\; \times
{2 \alpha_s(\mu^2) \alpha^2 e^2_c \pi^2 \over m_c (Q^{2}+(2m_{c})^{2})}
\Bigg\{ {1+(1-y)^2 \over y}
\Big[ \langle{\cal O}^{\psi}_8(^1S_0)\rangle +
{3Q^2+7(2m_c)^2 \over Q^{2}+(2m_{c})^{2}}
{\langle{\cal O}^{\psi}_8(^3P_0)\rangle \over m^2_c} \Big]
\nonumber \\
& & \;\;\;\;\;\;\;\;\;\;\;\;\;\;
- y {8(2m_c)^2Q^2 \over (Q^{2}+(2m_{c})^{2})^2}
{\langle{\cal O}^{\psi}_8(^3P_0)\rangle \over m^2_c} \Bigg\} \; ,
\label{epcs}
\end{eqnarray}
where $s$ is the electron-proton center-of-mass energy squared, 
$\mu^2= Q^2 + (2m_c)^{2}$ is both the factorization and renormalization
scale, and $f_{g/p}(x)$ is the gluon distribution function
of the proton. The momentum fraction of the
virtual photon relative to the incoming lepton is
$y \equiv P_{p}\cdot q / P_{p} \cdot k$, where $P_{p}$ is the proton
four-momentum, $q$ is the photon four-momentum, and $k$ is the
incoming lepton four-momentum, and $Q^{2} \equiv -q^{2}$.
Note that the relative importance of
$\langle {\cal O}^{\psi}_8(^1S_0)\rangle$
and $\langle {\cal O}^{\psi}_8(^3P_0)\rangle$
changes as a function of $Q^2$. Thus it is possible to fit the
differential cross section as a function of $Q^2$ 
and extract both of these matrix elements.

The result presented in Eq.~(\ref{epcs}) holds
for all values of $Q^{2}$. Taking the limit $Q^2 \to 0$ one recovers
the photoproduction cross section~\cite{Photo} convoluted with the electron 
splitting function:
\begin{equation}
\lim_{Q^2\to 0}\sigma(e+P\to e+\psi+X)\to {\alpha \over 2 \pi}
\int{dQ^2 \over Q^2} \; \int^1_0 dy \; {1+(1-y)^2\over y}
\hat{\sigma}(\gamma + P \to \psi + X) \; .
\label{Qgoto0}
\end{equation}
As mentioned in the introduction, corrections to the photoproduction 
cross section from higher order perturbative QCD corrections terms,
from diffractive photoproduction, and from higher twist effects, may
all be large.
However, in the high-energy limit $Q^2, s \gg (2m_c)^2$ we expect
corrections to be negligible. Letting
{} $Q^2, s \gg (2m_c)^2$ in Eq.~(\ref{epcs}) we obtain
\begin{eqnarray}
\lefteqn{
\lim_{m^2_c/Q^2,m^2_c/s\to 0}\sigma(e+P\to e+ \psi+X)\to 
\int{dQ^2 \over Q^2} \; \int dy \; \int dx \; f_{g/N}(x) 
\; \delta(xys-Q^2) }
\nonumber \\
& & \;\;\;\;\;\;\;\; \times
{2 \alpha_s(Q^2)\alpha^2 e^2_c \pi^2 \over m_c
Q^2}\;{1+(1-y)^2\over y}
\left( \langle{\cal O}^{\psi}_8(^1S_0)\rangle + 3
{\langle{\cal O}^{\psi}_8(^3P_0)\rangle \over m^2_c} \right)
\; .
\label{helim}
\end{eqnarray}

Once the color-octet matrix elements have been extracted from
measurements of the total cross section, they can then be used to
make predictions for the polarization of $J/\psi$ produced in
leptoproduction without introducing
any new free parameters. The measurement of the polarization will be 
an important check of the NRQCD factorization formalism.
The cross section for production of
longitudinally polarized $J/\psi$, where the polarization axis is  
the direction
of the three-momentum of the $J/\psi$ in the photon-proton
center-of-mass frame,
is:
\begin{eqnarray}
\label{Pol}
\lefteqn{
\sigma(e+p\to e+\psi_{L} +X) = \int {dQ^2 \over Q^2}\int dy
\int dx\; f_{g/p}(x)\; \delta(xys-(2m_c)^{2}-Q^2) }
\nonumber \\
& & \;\;\;\;\;\;\;\;\; \times
{2 \alpha_s(\mu^2) \alpha^2 e^2_c \pi^2 \over 3 m_c(Q^{2}+(2m_{c})^{2})}
\Bigg\{ {1+(1-y)^2 \over y}
\Big[ \langle{\cal O}^{\psi}_8(^1S_0)\rangle +
3 {\langle{\cal O}^{\psi}_8(^3P_0)\rangle \over m^2_c} \Big]
\nonumber \\
& & \;\;\;\;\;\;\;\;\;\;\;\;\;\;\;\;\;\;
 + {1-y \over y}
{48 (2m_c)^2 Q^2 \over (Q^{2}+(2m_{c})^{2})^{2}}
{\langle{\cal O}^{\psi}_8(^3P_0)\rangle \over m^2_c} \Bigg\} \; .
\end{eqnarray}

The polarization can be measured by studying the angular
distribution of the leptons in the leptonic decay of the $J/\psi$. If 
$\theta$ is defined to be the angle between the momentum of the
leptons in the $J/\psi$ rest frame and the momentum of the $J/\psi$ 
in the photon-proton center of momentum frame, then the decay
distribution of the leptons is given by:
\begin{equation}
{d \Gamma(\psi \rightarrow \ell^+ \ell^-) \over d {\rm cos} \theta} 
\propto 1 + \alpha~{\rm cos}^2 \theta \nonumber
\end{equation}
where,
\begin{equation}
\label{polparm}
\alpha = {1 - 3 f_L \over 1 + f_L},~~~~~f_L \equiv {\sigma(\psi_L) \over 
\sigma(\psi)} \; .
\end{equation}
Note that the
theoretical prediction depends only on one parameter, the ratio
$R \equiv \langle {\cal O}^{\psi}_8(^3P_0) \rangle /(m_c^2
\langle {\cal O}^{\psi}_8(^1S_0) \rangle)$. In the
limit where $Q^2, s \gg (2m_c)^2 $, the polarization parameter,
$\alpha$, goes to zero. This can be seen by inspection of
Eqs.~(\ref{helim}) and~(\ref{Pol}).

\begin{figure}
\epsfxsize=7cm
\hfil\epsfbox{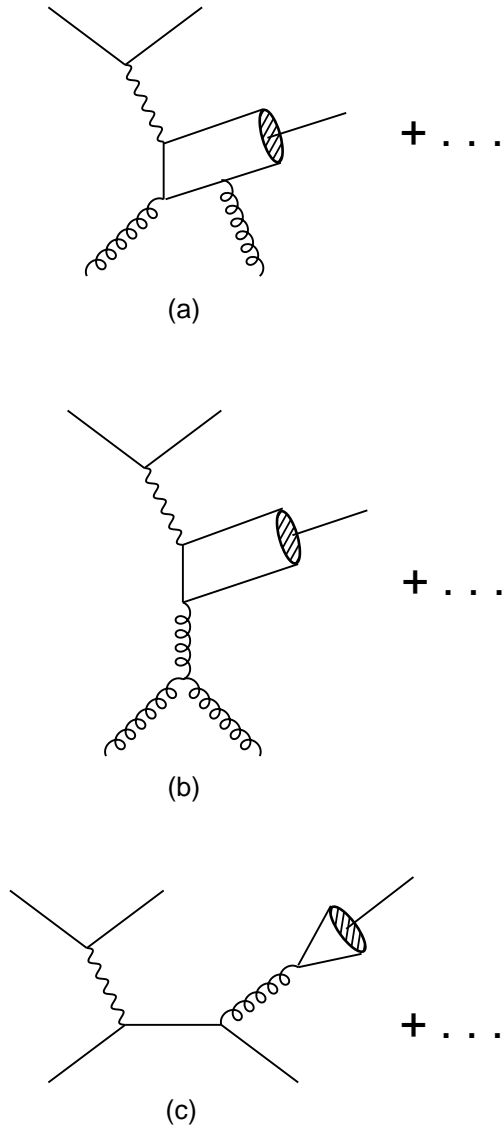}\hfill
\caption{ $O(\alpha_s^2)$ diagrams for leptoproduction
of $J/\psi$. There are six diagrams of the type shown in (a), 
two diagrams of the type shown in (b), and two diagrams of the type
shown in (c). The remaining quark diagrams, which are not shown, can
be obtained from the diagrams in (b) by replacing the external gluon
lines with quarks. Note, that only the diagrams in (a) 
contribute to the production of a $c\bar{c}$ pair in a color-singlet
${}^3S_1$ configuration.}
\label{nlo}
\end{figure}

At $O(\alpha_{s}^2)$, $J/\psi$ is produced 
through the hadronization of a $c\bar{c}$ pair in either a  
${}^{3}S_{1}$,
${}^{1}S_{0}$, or ${}^{3}P_{J}$ configuration . The Feynman diagrams 
are shown in Fig.~\ref{nlo}.
Note that only the diagrams in Fig.~\ref{nlo}a produce
a $c\bar{c}$ pair in a color-singlet ${}^{3}S_1$ state; all other
diagrams produce a color-octet $c\bar{c}$ pair.

Once again we use the techniques of Ref.~\cite{BC} to compute the
$J/\psi$ production cross section from the Feynman diagrams shown in 
Fig~\ref{nlo}. The
expressions obtained are complicated, so we do not
present them here\footnote{The FORTRAN code for generating the
differential cross sections presented later in this work is
available by request.}. However, we do report on checks we
have made to ensure that that these expressions are correct.  First, we
have checked that in the $Q^{2} \to 0$ limit we recover the
photoproduction cross section convoluted with the electron
splitting function. Second, we
numerically compared our color-singlet result to a
calculation carried out by Merabet, Mathiot, and
Mendez-Galain~\cite{mmm}, and find agreement.

\section{Non-perturbative and Diffractive contributions}

Before presenting our predictions, we wish to discuss
non-perturbative corrections to the NRQCD factorization formalism.  
Precise determination of the NRQCD matrix elements will be
impossible if these effects are not controlled. In the first part  
of this section, we consider diffractive leptoproduction and higher  
twist corrections to the parton model. In the second part, we
discuss the breakdown of the NRQCD velocity expansion near the
boundaries of phase space.

\subsection{Diffraction and higher twist}

At the HERA collider at DESY, diffractive
processes contribute roughly 40\% of the total $\gamma-p$ cross
section \cite{HERA}, and may be an equally important contribution to 
leptoproduction of $J/\psi$. Moreover, the $J/\psi$
produced via  this mechanism have similar kinematics to $J/\psi$
produced via the $O(\alpha_s)$ color-octet contribution.

To understand why diffractive
leptoproduction of $J/\psi$ is kinematically similar to leading order 
color-octet production, it is necessary to introduce the following
variable:
\begin{equation}
\label{z}
z \equiv {P_{p} \cdot P_{\psi} \over P_{p}\cdot q} \; .
\end{equation}
Here $P_{\psi}$ is the four-momentum of the $J/\psi$. In the proton 
rest frame $z = E_{\psi}/E_{\gamma}$. For the process
$e + p \rightarrow e + J/\psi + X$, $z = 1 + t/(s + Q^2)$,
where we neglect the proton mass, and the mass of the final state
system $X$.  Diffractive processes are exponentially suppressed away from 
$t = 0$, where $z =1$.

The short-distance part of the $O(\alpha_s)$ color-octet contribution leads to the production of a $c\bar{c}$ pair with no other particles in the final 
state; hence one would expect $z=1$. However, soft gluons with
momentum of order $m_{c} v^{2}$ are radiated during the
non-perturbative evolution of the $c\bar{c}$ into the $J/\psi$,
resulting in $z \approx 1-v^{2}$. Thus there is significant
overlap of the kinematic regimes of the $O(\alpha_s)$ color-octet contribution
and the diffractive contribution. It is possible to eliminate the
diffractive contribution by requiring $z \ll 1$; however the $O(\alpha_s)$
color-octet mechanism will also be eliminated by such a cut. There are 
other color-octet contributions at $z \ll 1$, but they only occur 
at higher orders in $\alpha_{s}$ where there is no longer a
perturbative enhancement of the color-octet term relative to the
color-singlet term.

Because diffractive leptoproduction is characterized by a large
scale, $Q^2$, this process can be studied using perturbative QCD. In  
Ref. \cite{Diffractive}, a perturbative analysis of diffractive  
leptoproduction predicts the cross section to fall as $1/(Q^2 + (2 m_c)^2)^3$, as compared to $1/(Q^2 + (2 m_c)^2)^2$ for the color-octet mechanism. Therefore, at sufficiently large $Q^2$, the diffractive contribution should be a negligible correction to our calculation. Furthermore,
Ref.\cite{Diffractive} predicts that diffractively produced
$J/\psi$ will be longitudinally polarized in the limit $Q^2 >> (2
m_c)^2$. As discussed earlier in this paper, the $O(\alpha_s)$ color-octet
mechanism predicts leptoproduced $J/\psi$ to be unpolarized in this  
limit. Therefore, polarization of the $J/\psi$ will be  a useful
tool for distinguishing diffractive production from color-octet
mechanisms.

There are also obvious kinematic differences between
color-octet and diffractive production which should make it easy to distinguish between the two. The most important distinction is that color-octet production will {\it not} lead to a rapidity gap in the final state, the hallmark signature of diffractive production. In color-octet production, the octet
$c \bar{c}$ pair produced in the short-distance process must
radiate at least one soft gluon before hadronizing into the final
state $J/\psi$. This soft gluon, with energy in the quarkonium rest 
frame of order $m_c v^2$, will be well-separated in phase space
from  the proton remnant. A simple computation shows that the
invariant mass of the soft gluon emitted by the color-octet
$c\overline{c}$ combined with the proton remnant will be
approximately $v W_{\gamma p}$, where $W_{\gamma p}$ is the
center-of-mass system energy of the virtual photon and proton. For the
H1 experiment at HERA, $W_{\gamma p}$ ranges between $30~{\rm GeV}$
and $150~{\rm GeV}$~\cite{H1}. If we take $v^2 = 0.25$,  then for the
color-octet contribution to $e + p \rightarrow e + J/\psi + X$, we
expect $M_X \geq 15~{\rm GeV}$.  
This is obviously much greater than $M_X = 1~{\rm GeV}$, what is
expected from elastic diffractive events. For diffractive
disassociative events, $M_X$ can be greater than 1 GeV, but the cross 
section is expected to fall off as $1/M_X^2$ \cite{H1}. By requiring 
that there be no rapidity gap in the event and that $M_X \gg 1 {\rm 
GeV}$, it should be possible to extract a clean signal for
color-octet $J/\psi$ production.

In addition to the diffractive contribution, we must also consider the 
possible contribution from higher twist corrections to the
parton model. These higher twist effects can be of
great importance in quarkonium production calculations.

Well-known factorization theorems \cite{Factor} show that the inclusive 
production of a hadron can be written as a convolution of a parton
distribution function, a parton scattering cross section calculable 
in perturbative QCD, and a fragmentation function which describes how the
parton produced in the short distance process evolves into the final 
state hadron. This factorization formula receives higher twist
corrections that scale as $(\Lambda_{QCD}/\mu)^n$, where $\mu$ is a 
characteristic scale of the process. For inclusive single hadron
production, $\mu$ is $P_{\perp}$, the momentum of the hadron
transverse to the beam axis. By requiring that $P_{\perp}$ be large 
compared to $\Lambda_{QCD}$, we can ensure that predictions of the
parton model calculation will be accurate. For quarkonium production, 
the situation is more complicated because there are two large scales 
that characterize the process: the mass of the quarkonium state,
$M_H \approx 2 m_Q$,  and $P_{\perp}$.  If some higher twist corrections 
are suppressed by $M_H$ rather than $P_{\perp}$, nonperturbative
corrections to the parton model will not diminish as $P_{\perp}$
increases. This problem is particularly worrisome in the case of
charmonium production, since $2 m_c \approx 3~\rm{GeV}$, so
nonperturbative corrections are not strongly suppressed.

Higher twist corrections to color-singlet {\it photoproduction} are 
computed by Ma in Ref.~\cite{Twist}. There it is shown that some of the 
twist-4 corrections are suppressed only by $M_{\psi}^2$. Since higher 
twist structure functions are unknown, it is impossible to calculate 
the size of corrections, but it is clear that, at energies typical  
of the HERA collider, $\sqrt{s_{\gamma p}}\approx 100~{\rm GeV}$,
they can be important. Ma states in
Ref.~\cite{Twist} that in the case of electroproduction for fixed $Q^2$ there are no twist-4 corrections that are suppressed by only $M_{\psi}^2$. However, no detailed calculation is given, and until such a calculation is carried out it is not clear what kinematical factor suppresses the higher twist
contributions.  We assume that higher twist corrections scale like $\Lambda_{QCD}^2/(Q^2 + (2 m_c)^2)$.  An explicit calculation would be most welcome.

\subsection{Leptoproduction shape functions}

Another class of nonperturbative effects is associated with higher
orders in the NRQCD velocity expansion \cite{Shape}.  For quarkonium 
production near the boundaries of phase space, it is sometimes the
case that contributions from NRQCD operators which are higher order in 
$v$ are enhanced by kinematical factors.  This results in
the breakdown of the NRQCD expansion.  The crux of the problem is
that in the perturbative QCD part of the matching calculation one uses 
twice the heavy quark mass instead of the quarkonium mass to  
compute the phase
space for the production of the quarkonium meson. The difference
between $2m_{Q}$ and $M_{H}$ is a $v^{2}$ correction, which is ignored
in leading-order calculations. However, at the boundaries of
phase space this difference becomes important, and it is necessary  
to sum an
infinite number of
NRQCD matrix elements.  This resummation leads to a universal
distribution function called a shape function.  Because of the
universality of the shape functions, it may be possible in the future 
to test NRQCD by comparing shape functions extracted from different 
quarkonium production processes.  However, in the present paper we are 
interested in extracting numerical values for the leading color-octet 
matrix elements.  Therefore, we want to make sure that the effects of 
higher order terms in the NRQCD expansion are genuinely suppressed by 
powers of $v^2$ and can be safely neglected.

The leading order contribution for producing $J/\psi$  through the
hadronization of a color-octet $c\bar{c}$ pair in a ${}^{1}S_{0}$
configuration is:
\begin{eqnarray}
\label{LO}
{d\sigma \over d Q^2 dy}&=& \int dx f_{g/P}(x)
{\sum|{\cal M}|^2 \over 16 \pi x s } {d^3 P_{\psi} \over 2 E_{\psi}} 
\delta^{(4)}(f + g - f^{\prime} - P_{\psi})  \\
&=& \int dx S(x,Q^2) {1 + (1-y)^2 \over y} \delta \left(y - {Q^2 + (2 
m_c)^2 \over x s}\right)
\nonumber \\
& & \;\;\;\;\;\;\;\;\;\;\;\; \times
\sum_X \langle 0 |\psi^{\dagger}T^a\chi|J/\psi+X\rangle \langle
J/\psi+X|\chi^{\dagger}T^a\psi|0\rangle \; , \nonumber
\end{eqnarray}
where
\[
S(x,Q^2) = {8 \pi f_{g/P}(x)  \alpha_s \alpha^2 e_c^2 \over 9 m_c
x s Q^2 (Q^2 +(2 m_c)^2)} \; .
\]
The ${\cal O}^{\psi}_8(^3P_0)$ term can be analyzed in an analogous  
manner. In
Eq.~(\ref{LO}), $P_{\psi} = p_c + p_{\bar{c}}$, and in evaluating the 
phase space integral we have used the approximation
$P_{\psi} = (2 m_c,0,0,0)$ in the quarkonium rest frame.
If we allow for the $c$ and $\bar{c}$ produced in the short distance 
process to have nonvanishing velocity in the quarkonium rest frame, 
Eq.~(\ref{LO}) becomes:
\begin{eqnarray}
\label{Correction}
{d\sigma \over d Q^2 dy} &=& \int dx f_{g/P}(x)
{\sum|{\cal M}|^2 \over 16 \pi x s } {d^3 P_{\psi} \over 2 E_{\psi}}
\delta^{(4)}(f + g - f^{\prime} - P_{\psi} - \Lambda(l_c + l_{\bar{c}})) 
\\
&=& \int dx S(x,Q^2){1 + (1-y)^2 \over y}
\delta \left(y - {Q^2 + (2 m_c)^2\over x s} -
{2 P_{\psi}\cdot\Lambda(l_c + l_{\bar{c}}) \over x s} \right)
\nonumber \\
& & \;\;\;\;\;\;
\times \sum_X \langle 0 |\psi^{\dagger}T^a\chi|J/\psi+X\rangle \langle 
J/\psi+X|\chi^{\dagger}T^a\psi|0\rangle \; . \nonumber
\end{eqnarray}
In this expression,
$l_{c(\bar{c})} =
(\sqrt{\vec{l}_{c(\overline{c})}^2 + m_c^2} -
m_c,\vec{l}_{c(\overline{c})})$ is the charm (anti-charm) quark
four-momentum in the $J/\psi$ rest frame,
and $\Lambda$ is a boost that takes us from the quarkonium rest frame to 
the frame in which we are performing the calculation. In order to  
perform the
matching onto NRQCD, the delta function in Eq.~(\ref{Correction})  
is expanded
in powers of $2 P_{\psi} \cdot\Lambda(l_c + l_{\bar{c}})$ and these 
factors are identified with the matrix elements of the NRQCD operators:
\begin{eqnarray}
\label{corr2}
\lefteqn{
{d\sigma \over d Q^2 dy} = \int dx S(x,Q^2){1 + (1-y)^2 \over y}
\sum_n {1 \over n!} \delta^{(n)}\left(y - {Q^2 + (2 m_c)^2 \over x s} 
\right) } \\
& & \;\;\;\;\;\;\;
\times \sum_X \langle 0 |\psi^{\dagger}T^a\chi|J/\psi +X\rangle  
\langle
J/\psi +X|\left({8 m_c^2 i\hat{D}_0\over x s}\right)^n
\chi^{\dagger}T^a\psi|0\rangle \; .\nonumber
\end{eqnarray}
Here, $\hat{D}_0 = D_0/(2 m_c)$, where $D_0$ is the time component  
of the
gauge covariant derivative. The infinite series of operators in
Eq.~(\ref{corr2}) can be
summed into the universal shape function:
\begin{equation}
\label{CSwSF}
{d\sigma \over d Q^2 dy} = \int dx \int dy_E S(x,Q^2){1 + (1-y)^2  
\over y}
\delta\left(
y - {Q^2 + (2 m_c)^2 \over x s } - {8 m_c^2 \over x s} y_E
\right)
F[^1S_0^{(8)}](y_E),
\end{equation}
where
\begin{equation}
\label{ShapeFunction}
F[^1S_0^{(8)}](y_E) = \sum_X \langle 0
|\psi^{\dagger}T^a\chi|H+X\rangle \langle H+X|\delta\left({y_E
-i\hat{D}_0}\right) \chi^{\dagger}T^a\psi|0\rangle.
\end{equation}

We now wish to consider the singly differential distribution
$d \sigma /d Q^2$. Integrating over $y$ in Eq.~(\ref{CSwSF}), and
expanding in powers of $y_E$ we find:
\begin{eqnarray}
{d\sigma \over d Q^2}&=& \int dx \int dy  \int dy_E \; S(x,Q^2)
{1 + (1-y)^2 \over y} F[^1S_0^{(8)}](y_E)
\; \delta\left(
y - {Q^2 + (2 m_c)^2 \over x s } - {8 m_c^2 \over x s} y_E
\right)
\nonumber \\
&=&\int dx \; S(x,Q^2)
\left(
{1 + (1-y_{LO})^2 \over y_{LO}} \langle {\cal O}_8(^1S_0)\rangle -
{2-y_{LO} \over y_{LO}} \Delta F^{(1)} +
{2 \over y_{LO}} \sum_{n=2}^{\infty}(-)^n \Delta^n F^{(n)}
\right) .
\nonumber  \\
\mbox{}
\label{bae}
\end{eqnarray}
In this expression, $y_{LO} = (Q^2 + (2 m_c)^2)/x s$, $\Delta = 2 (2 
m_c)^2/(Q^2 + (2 m_c)^2)$, and
\[
F^{(n)} = \int d y_E~y_E^n F[^1S_0^{(8)}](y_E) \sim v^{2n}.
\]
Notice that for $Q^2 \gg (2m_c)^2$, $\Delta \ll 1$ and the higher
order corrections are suppressed by powers of $\Delta v^2$ instead of 
$v^2$. Thus the corrections associated with non-relativistic treatment 
of the quarkonium phase space can be made negligible by considering the 
singly differential distribution $d \sigma /d Q^2$ at large $Q^2$.

Higher order NRQCD corrections associated with the shape function can 
also be computed for the next-to-leading order color-octet and
color-singlet contributions to leptoproduction of $J/\psi$. The
calculation is nearly identical to the calculation for
photoproduction performed in Ref.~\cite{Shape}, so we will merely
quote results. The cross section is written in terms of the following 
variables: $Q^2$, $y$, $\hat{z}$, $P_{\perp}$, and $\phi$.  $\hat{z}$ 
is defined to be $P_p\cdot P_{\psi}/P_p \cdot q$, where the  
non-relativistic
approximation for $P_{\psi}$ is used ($P_{\psi} = (2 m_c,0,0,0)$ in the $J/\psi$ restframe). This variable is distinct from 
experimentally observed $z = P_p\cdot P_{\psi}/P_p \cdot q$, where
$P_{\psi}$ is the observed $J/\psi$ momentum.  We
define $P_{\perp}$ to be the momentum of the $J/\psi$ relative to the 
axis defined by the $\gamma^*$ three momentum in the $\gamma^*$-proton
center-of-momentum frame, and $\phi$ is the azimuthal angle of the
$J/\psi$ about this axis. The differential cross section is given by:
\begin{eqnarray}
\label{NLO}
\lefteqn{
{d \sigma \over dQ^2 dy d\hat{z}} =  \int d\phi \int dP_{\perp}^2
\int dy_+ \int dx
{f_{g/P}(x)\sum |{\cal M}|^2 \over 2^9 \pi^4 \hat{z} x s}  \;
 F[^{2S+1}L_J^{(1,8)}](y_+) }
 \\
& & \;\;\;\;\;\;
\times \delta \left( (x y s - Q^2)(1 - \hat{z}) -
{P_{\perp}^2 + (2 m_c)^2(1-\hat{z}) \over \hat{z}} -
{P_{\perp}^2 + (2 m_c)^2(1-\hat{z})^2 \over (2 m_c) \hat{z}(1 -  
\hat{z})}y_+
\right)
\nonumber
\end{eqnarray}
Eq.~(\ref{NLO}) applies equally well to color-octet and color-singlet 
production, so the quantum numbers of the $c\overline{c}$ produced in 
the short-distance process are not specified. It is worth noting that 
the shape function that appears in Eq.~(\ref{NLO}) differs from the 
shape function in Eq.~(\ref{ShapeFunction}): $\delta(y_E -
i\hat{D}_0)$ is replaced with $\delta(y_+ - in\cdot\hat{D})$, where 
$n^{\mu}$ is a light-like four vector. Expanding the delta function 
in Eq.~(\ref{NLO}) in powers of $y_+$, it is easy to see that the
moments of the shape function, $F^{(n)} = \int d y_+~y_+^n F[^{2S+1}L_J^{(1,8)}](y_+)$, are multiplied by a factor $(1-\hat{z})^{-n}$. For $1 - \hat{z} \leq v^2$, an infinite number of NRQCD matrix elements are relevant. For this reason, the differential distribution in $z$ cannot be calculated in the forward region, even if $Q^2 \gg 4 m_c^2$. However, other differential distributions, such as $d \sigma/d Q^2$ or $d \sigma/dP_{\perp}^2$, can be computed because the integration over $z$ provides sufficient smearing over the singular region.
 
Based on the analysis presented in this section we have reason to
believe that non-perturbative corrections are under 
control for leptoproduction of $J/\psi$ at large $Q^2$.  First, 
though the size of the diffractive
contribution to leptoproduction is difficult to estimate because
perturbative QCD calculations of this process are not expected to
correctly yield the absolute normalization of the cross section
\cite{Diffractive}, perturbative
QCD calculations do predict that diffractive production will be less
significant at large $Q^2$. In addition, rapidity gaps, the invariant
mass squared 
of the final hadronic state, and the polarization of the $J/\psi$
produced in diffractive production, all differ in a qualitative way
from color-octet production. These observables should allow
experimentalists to isolate a clean color-octet signal. 
Next, higher twist
effects are expected to scale as $\Lambda_{QCD}^2/(Q^2 + (2 m_c)^2)$.
Finally, examination of higher order $v^2$
corrections associated with shape functions gives us insight into
which differential distributions can be reliably computed within the
NRQCD formalism.  The distribution in $z$ near the kinematic endpoint
certainly cannot be predicted in the leading order approximation we
consider in this paper.  Therefore, we will not attempt to calculate
$d \sigma /d z$, even though this quantity is often measured in
experiments and has been considered by theorists in previous
studies. The quantities which we do compute should be insensitive to
these corrections.

\section{Matrix element phenomenology}

As discussed in the introduction, a naive counting of powers of $v$ and
$\alpha_s$ tells us that the leading terms in the NRQCD factorization
formula for the $J/\psi$ leptoproduction cross section are the $O(\alpha_{s})$ color-octet contribution, Eq.~(\ref{epcs}), and the
$O(\alpha_{s}^2)$ color-singlet contribution. The
color-octet contribution depends on the NRQCD production matrix
elements $\langle{\cal O}^\psi_8({}^1S_0) \rangle$ and
$\langle{\cal O}^\psi_8({}^3P_0) \rangle$, and the color-singlet
contribution depends on $\langle{\cal O}^\psi_1({}^3S_1) \rangle$.
These production matrix elements must be treated as parameters.
Thus, before we present our prediction for $J/\psi$ leptoproduction we discuss the current status of NRQCD $J/\psi$ production phenomenology.

The color-singlet matrix element can be determined from the decay
$J/\psi \to \mu^{+} \mu^{-}$ or from lattice calculations~\cite{bsk}. 
The value determined for
$\langle {\cal O}^\psi_1({}^3S_1)\rangle$ is shown in Table~\ref{COME}.

The values of the color-octet matrix elements
$\langle {\cal O}^\psi_8({}^1S_0)\rangle$,
$\langle {\cal O}^\psi_8({}^3P_0)\rangle$, and
$\langle {\cal O}^\psi_8({}^3S_1)\rangle$
have been extracted from CDF measurements~\cite{CDF}
by Cho and Leibovich \cite{CL}, and by
Beneke and Kr$\ddot{\rm{a}}$mer \cite{BK}.  Neither analysis
extracts $\langle {\cal O}^{\psi}_8(^1S_0)\rangle$ and
$\langle {\cal O}^{\psi}_8(^3P_0)\rangle$ independently, instead  
they fit a
linear combination,
$M_{k} = \langle {\cal O}_8(^1S_0)\rangle +
k~\langle {\cal O}_8(^3P_0)\rangle/m_c^2$, of the two. In
Ref.~\cite{CL} $k=3$, and in Ref.~\cite{BK} $k = 3.5$. A detailed
discussion of theoretical uncertainties in the extraction of these
matrix elements is given in Ref.~\cite{BK}, and the results of their 
fit are shown in Table~\ref{COME}.  The largest uncertainty is due to 
scale setting ambiguities.  Statistical errors are relatively small, 
only $\sim 15 \% $ for $\langle {\cal O}_8(^3S_1) \rangle$, and $\sim 
25\% $ for $M_{3.5}$.  In this analysis, the charm quark mass is
chosen to be 1.5 GeV and errors associated with uncertainty in the
choice of this parameter are not considered.  Later we will consider errors in our calculation of leptoproduction of $J/\psi$, and we will find that uncertainty in the charm quark mass is a much greater source of error than uncertainty due to scale setting ambiguities. While this may not be the case in hadroproduction, we nevertheless feel that uncertainty in charm quark mass is a potentially large source of error which has not been included in the analysis of Ref.~\cite{BK}. This additional source of error must be kept in mind when considering the extraction of NRQCD matrix elements from this process.
\begin{table}
\begin{tabular}{cccc}
  & $\langle {\cal O}^\psi_1({}^3S_1)\rangle \; ({\rm GeV}^3) $
& $\langle {\cal O}^\psi_8({}^3S_1)\rangle \; (10^{-2}~{\rm GeV}^3)$
& ${\rm M_{3.5}} \; (10^{-2}~{\rm GeV}^3)$ \\
\hline
$\Gamma (J/\psi \to \mu^+ \mu^-)$
& 1.1$\pm 0.1$
& --
& -- \\
CTEQ4L
& --
& 1.06$\pm 0.14^{+1.05}_{-0.59}$
& 4.38$\pm 1.15^{+1.52}_{-0.74}$ \\
GRV(1994)LO
& --
& 1.12$\pm 0.14^{+0.99}_{-0.56}$
& 3.90$\pm 1.14^{+1.46}_{-1.07}$ \\
MRS(R2)
& --
& 1.40$\pm 0.22^{+1.35}_{-0.79}$
& 10.9$\pm 2.07^{+2.79}_{-1.26}$ \\
\end{tabular}
\caption{$M_{3.5}$ is the
linear combination
$\langle O_8(^1S_0) \rangle + 3.5 \langle O_8(^3P_0) \rangle /m_c^2$. 
The color-singlet matrix element is determined from the decay rate
for $J/\psi \to \mu^{+} \mu^{-}$. The color-octet matrix elements are 
determined from a fit to CDF data performed in
Ref. \protect\cite{BK}. First error is statistical, second is due to
scale uncertainty.}
\label{COME}
\end{table}

The extraction of $M_{3.5}$ appears to be particularly unreliable.
The contributions to the cross section proportional to the matrix  
elements
$\langle {\cal O}^{\psi}_8(^1S_0) \rangle$ and
$\langle {\cal O}^{\psi}_8(^3P_0) \rangle$ are subleading except at 
small transverse momentum, $P_{\perp}\leq 5~{\rm GeV}$. Thus the  
extraction
of these matrix elements is very sensitive to
effects which can modify the shape of the $P_{\perp}$ spectrum.
Specifically, the shape of the proton gluon density at small $x$
can modify the slope
of the $P_{\perp}$ distribution at low $P_{\perp}$.
This is reflected in the extreme sensitivity  of the value of the  
parameter
$M_{3.5}$ (see Table~\ref{COME}) to the choice of the
parton distribution function. The authors of \cite{BK} take this as 
evidence that the result of the fit for $M_{3.5}$ will be unstable to 
higher order corrections and nonperturbative effects that will modify 
the $P_{\perp}$ distribution at low $P_{\perp}$. As a result they
conclude that
$\langle {\cal O}^{\psi}_8(^1S_0) \rangle$ and
$\langle {\cal O}^{\psi}_8(^3P_0) \rangle$ are not
reliably extracted from CDF data.

As discussed in the introduction 
$\langle {\cal O}^{\psi}_8(^1S_0)\rangle$ and
$\langle {\cal O}^{\psi}_8(^3P_0) \rangle$ can be extracted from
experimental data on photoproduction and low energy hadroproduction of
$J/\psi$. However, theoretical uncertainties only allow for an order
of magnitude estimate of the color-octet matrix elements. 

\section{Results}

Our study of leptoproduction of $J/\psi$ at large $Q^2$ will show that
this process can provide a more accurate extraction of $\langle {\cal
O}^{\psi}_8(^1S_0) \rangle$ and $\langle {\cal O}^{\psi}_8(^3P_0)
\rangle$ than hadroproduction or photoproduction.  Unfortunately, at
this time, there does not exist appropriate data to which a reliable
fit can be made.  There does exist data on $J/\psi$
muoproduction~\cite{EMC}; a fit to the EMC data was carried out in
Ref.~\cite{SF}. However, we must caution that this data was taken with
low to moderate values of $Q^{2}$. In this regime, we expect large
non-perturbative corrections to the NRQCD factorization
formalism. More importantly, as we will see below, in the low $Q^2$
region the perturbative
calculation suffers errors, due to
the uncertainty in the charm quark mass parameter. Therefore, we will
not make a fit to the existing leptoproduction data. Instead, we will
choose two sets of reasonable values for $\langle {\cal
O}^{\psi}_8(^1S_0) \rangle$, and $\langle {\cal O}^{\psi}_8(^3P_0)
\rangle$ in order to make predictions.

In the results presented below,
 we choose either
$\langle {\cal O}^{\psi}_8(^1S_0) \rangle = 0.01 \;  
\mbox{GeV}^{3}$ and
$\langle {\cal O}^{\psi}_8(^3P_0) \rangle/m^{2}_{c} = 0.005 \;  
\mbox{GeV}^{3}$,
inspired by the order-of-magnitude estimates presented in
Table~\ref{COME}, and in accordance with NRQCD $v$-scaling rules, or
$\langle {\cal O}^{\psi}_8(^1S_0) \rangle = 0.04 \;  
\mbox{GeV}^{3}$, and
$\langle {\cal O}^{\psi}_8(^3P_0) \rangle /m^{2}_{c} = -0.003 \;
\mbox{GeV}^{3}$, which is consistent with the fit to 
low $Q^2$ EMC data carried out in Ref.~\cite{SF}.
In addition, we require $Q^{2} > 4 \; \mbox{GeV}^{2}$. This should reduce errors from higher order perturbative QCD terms and from higher twist effects. We also assume that experiments at HERA are most likely to make precision
measurements of $J/\psi$ leptoproduction, so we choose the
center-of-mass energy to be $\sqrt{s} = 300 \; \mbox{GeV}$, and
require $30 \; \mbox{GeV} < W_{\gamma p} < 150 \; \mbox{GeV}$, where 
$W^{2}_{\gamma p} = (P_{p}+q)^{2}$.

Given the above conditions, we calculate the $O(\alpha_s)$ color-octet
contribution to the $J/\psi$ leptoproduction cross section,
and the $O(\alpha_s^2)$ color-singlet contribution to the cross section. The
results are presented in Table~\ref{csres} for two sets of kinematic
cuts: $Q^2 > 4 \; \mbox{GeV}^{2}$, and $Q^{2} > 4 \; \mbox{GeV}^{2}$, 
$P^2_{\perp} > 4 \; \mbox{GeV}^{2}$, where $P_{\perp}$ is the
transverse momentum of the $J/\psi$ in the HERA lab frame. The
color-octet result is for the first (second) choice of 
$\langle {\cal O}^{\psi}_8(^1S_0) \rangle$ and
$\langle {\cal O}^{\psi}_8(^3P_0) \rangle$ given above. Note that with
no $P_{\perp}$ cut the color-singlet contribution is $\sim 1/4$ the
size of the color-octet contribution, while with a  $P_{\perp}$ cut it
is $\sim 1/5$ the size of the color-octet contribution. For either choice of color-octet matrix element values the $O(\alpha_s^2)$ color-singlet contribution is small compared to the $O(\alpha_s)$ color-octet contribution. 
\begin{table}
\begin{tabular}{cccc}
Cuts
& $O(\alpha_s)$ color-octet
& $O(\alpha_s^2)$ color-singlet
& $O(\alpha_s^2)$ color-octet \\
\hline
$Q^2 >4\;\mbox{GeV}^2$
& 331 (226) pb
& 89 pb
& -- \\
$Q^2, \; P^2_{\perp}>4\;\mbox{GeV}^2$ 
& 290 (200) pb
& 40 pb
& -- \\
$(P_{\perp}^{\gamma P})^2 > 2\;\mbox{GeV}^2, \; z < 0.8$
& --
& 13 pb
& 8 pb\\
\end{tabular}
\caption{The $O(\alpha_s)$ color-octet contribution to the
$J/\psi$ leptoproduction cross section for
$\langle {\cal O}^{\psi}_8(^1S_0) \rangle = 0.01 \; \mbox{GeV}^2$ and  
$\langle {\cal O}^{\psi}_8(^3P_0)\rangle /m^2_c=0.005\;\mbox{GeV}^2$
($\langle {\cal O}^{\psi}_8(^1S_0) \rangle = 0.04 \; \mbox{GeV}^2$ and, 
$\langle {\cal O}^{\psi}_8(^3P_0)\rangle /m^2_c=0.005\;\mbox{GeV}^2$),
the $O(\alpha_s^2)$ color-singlet contribution, and the $O(\alpha_s^2)$ color-octet contribution. The results are given for three different sets of
kinematic cuts. Only the $O(\alpha_s)$ color-octet calculation and the $O(\alpha_s^2)$ color-singlet calculation are valid in the region of the first two sets of cuts. There are only $O(\alpha_s^2)$ contributions in region of the third set of cuts. Note theoretical errors are not included. }
\label{csres}
\end{table}

If we consider final states in which a gluon jet is well separated
from the $J/\psi$, there is no contribution to the cross section from
the $O(\alpha_s)$ color-octet term.  This is because the $O(\alpha_s)$
color-octet contribution produces $J/\psi$ through the emission of
soft gluons from the $c\bar{c}$ pair produced in the hard scattering.
These gluons have less than $1 \; \mbox{GeV}$ of momentum in the
$J/\psi$ rest frame.  However, there will still be a color-octet
contribution from $O(\alpha_s^2)$ tree level diagrams.  To isolate
events in which the gluon jet is well separated from the $J/\psi$, we
require that $(P^{\gamma P}_{\perp})^2 > 2~{\rm GeV}^2$ and $z <
0.8$. $P^{\gamma P}_{\perp}$ is the momentum of the $J/\psi$
transverse to the axis defined by photon three-momentum in the
photon-proton center-of-momentum frame.  The result of this
calculation is shown in Table~\ref{csres}.  The $O(\alpha^2_s)$
color-octet diagrams contribute about 40\% of the total cross section.
We do not show differential distributions in $P^2_\perp$, rapidity, and
$Q^2$, but report that the shape of these distributions is roughly the same
for both $O(\alpha_s^2)$ color-octet and $O(\alpha_s^2)$ color-singlet
contributions. Therefore, inclusion of color-octet mechanisms only
serves to change the overall normalization of the total cross section
in this region of phase space. Since the overall normalization is
already uncertain in a leading order calculation, we do not regard
events with $J/\psi$ and a gluon jet as particularly interesting for
measuring color-octet matrix elements.

\begin{figure}
\rotate[l]{
\epsfxsize=10cm
\hfil\epsfbox{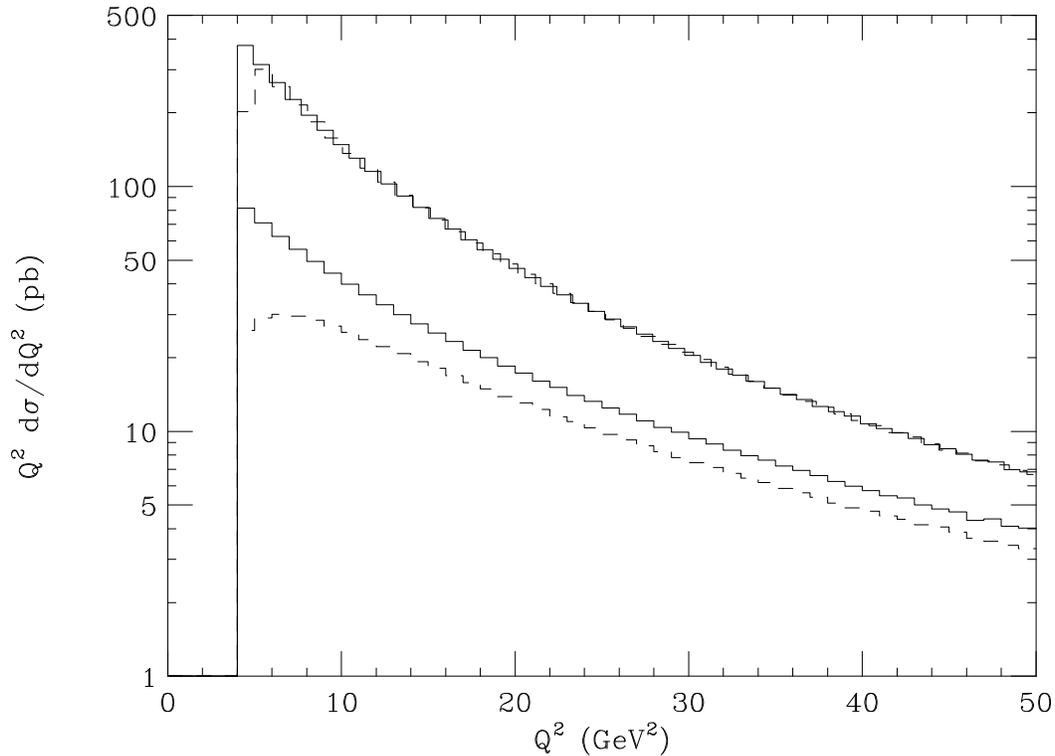}\hfill }
\caption{$O(\alpha_s)$ color-octet (upper histograms) and $O(\alpha_s^2)$ color-singlet (lower histograms) contributions to the differential cross section as a function of $Q^{2}$ for two sets of kinematic cuts. The solid histograms were generated with a $Q^2$ cut of $4 \; \mbox{GeV}^2$, and the dashed histograms were generated cutting on both $Q^2 > 4 \; \mbox{GeV}^2$, and 
$P^2_{\perp} > 4 \; \mbox{GeV}^2$. }
\label{lonlocomp}
\end{figure}

\begin{figure}
\rotate[l]{
\epsfxsize=10cm
\hfil\epsfbox{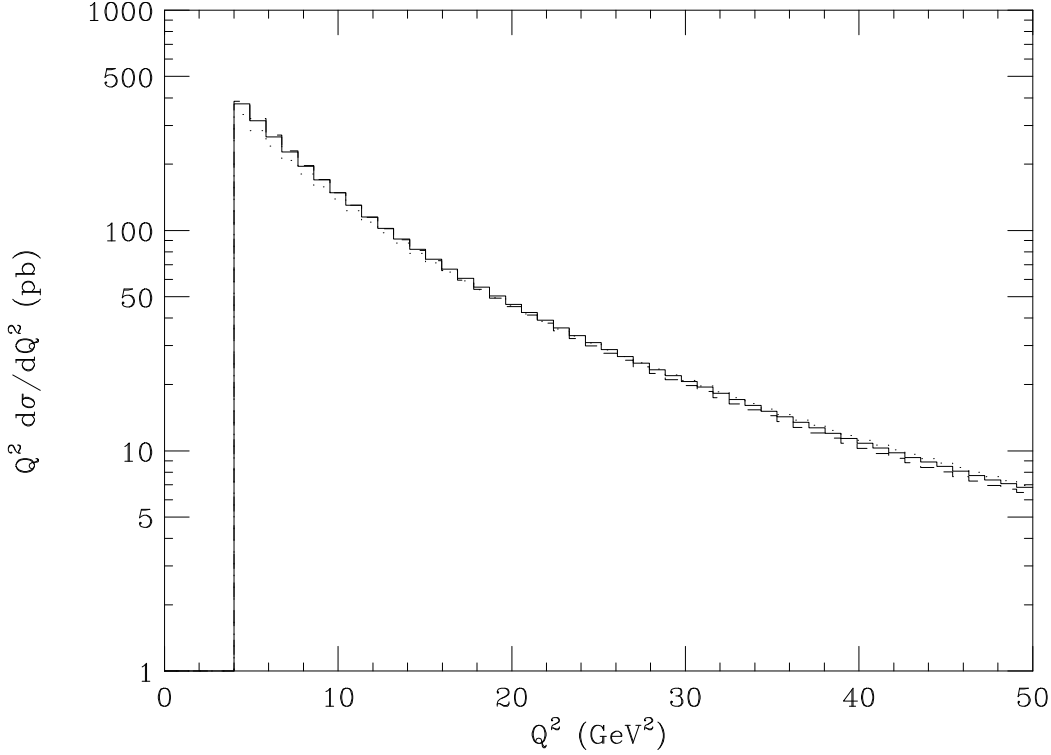}\hfill }
\caption{$O(\alpha_s)$ color-octet differential cross section as a function of 
$Q^{2}$ for three different choices of scale: $\mu^2 =
Q^{2}+(2m_{c})^{2}$ (solid histogram), $\mu^2/4$ (dotted histogram),
and $4 \mu$ (dashed histogram).}
\label{loQ}
\end{figure}

In Fig.~\ref{lonlocomp}, we show the $O(\alpha_s)$ color-octet
contribution (upper histograms) and the $O(\alpha_s^2)$ color-singlet
contribution (lower histograms) to the differential cross section as a
function of $Q^2$. The solid histograms are without a $P^2_{\perp}$ cut,
and the dashed histograms are with a 
$P^2_{\perp}$ cut of $4 \; \mbox{GeV}^2$. In generating the curves we
used $\mu^2 = Q^{2}+(2m_{c})^{2}$, $m_{c}=1.5 \; \mbox{GeV}$, and 
$\langle {\cal O}^{\psi}_8(^1S_0) \rangle = 0.01 \; \mbox{GeV}^3$,   
$\langle {\cal O}^{\psi}_8(^3P_0)\rangle /m^2_c=0.005\;\mbox{GeV}^3$.  
Next, we study the error in our calculation that results from uncertainty in the choice of renormalization and factorization scales. In Fig.~\ref{loQ}, we show the $O(\alpha_s)$ color-octet differential cross section as a function of
$Q^{2}$ for three choices of the renormalization 
and factorization scale: $\mu^2 = Q^{2}+(2m_{c})^{2}$ (solid histogram),  
$\mu^2/4$ (dotted histogram), and $4 \mu^2$ (dashed histogram). 
If we were to fit this result to
experimental data the scale uncertainty would
result in a theoretical uncertainty of about $^{+3\%}_{-11\%}$ at 
$Q^2 = 4~{\rm GeV}^2$, and $^{+2\%}_{-5\%}$ at $Q^2 = 50~{\rm GeV}^2$ in determining the values of the color-octet matrix elements. These small errors give us confidence that higher order perturbative corrections to the leading color-octet production mechanisms are under control.

\begin{figure}
\rotate[l]{
\epsfxsize=10cm
\hfil\epsfbox{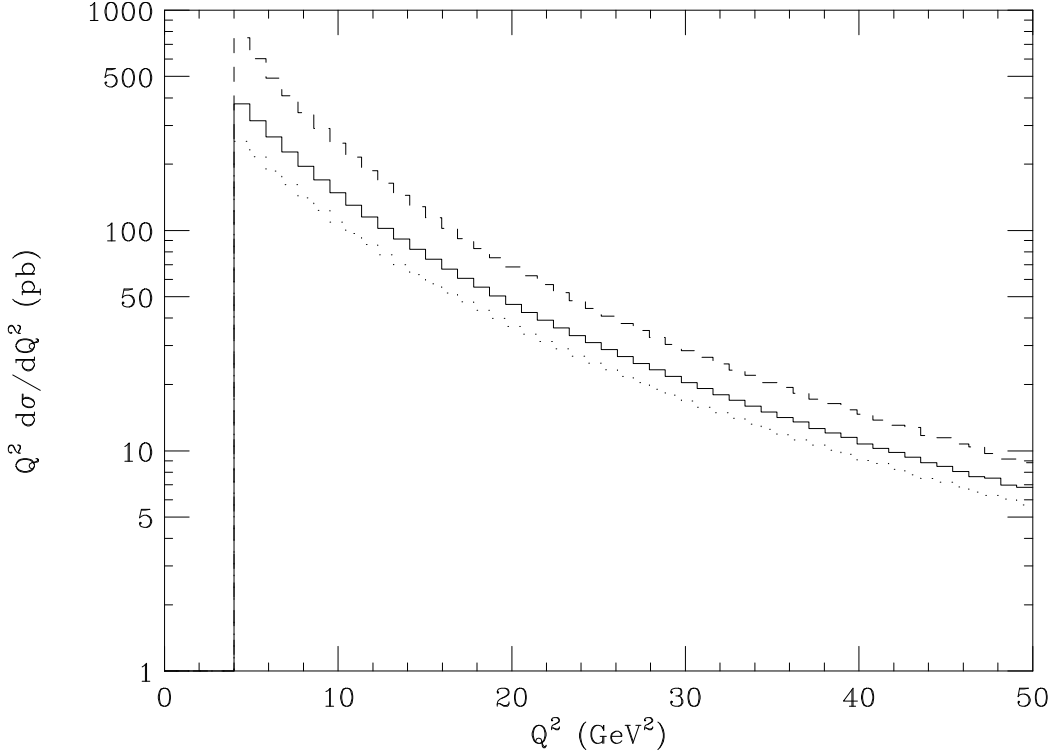}\hfill }
\caption{$O(\alpha_s)$ color-octet differential cross section as a function of 
$Q^{2}$ for three different choices of $m_{c}$. $m_{c}=1.3 \;
\mbox{GeV}$ (dotted histogram), $m_{c}=1.5 \; \mbox{GeV}$ (solid
histogram), and  $m_{c}=1.7 \; \mbox{GeV}$ (dashed histogram).}
\label{loQmc}
\end{figure}

\begin{figure}
\rotate[l]{
\epsfxsize=10cm
\hfil\epsfbox{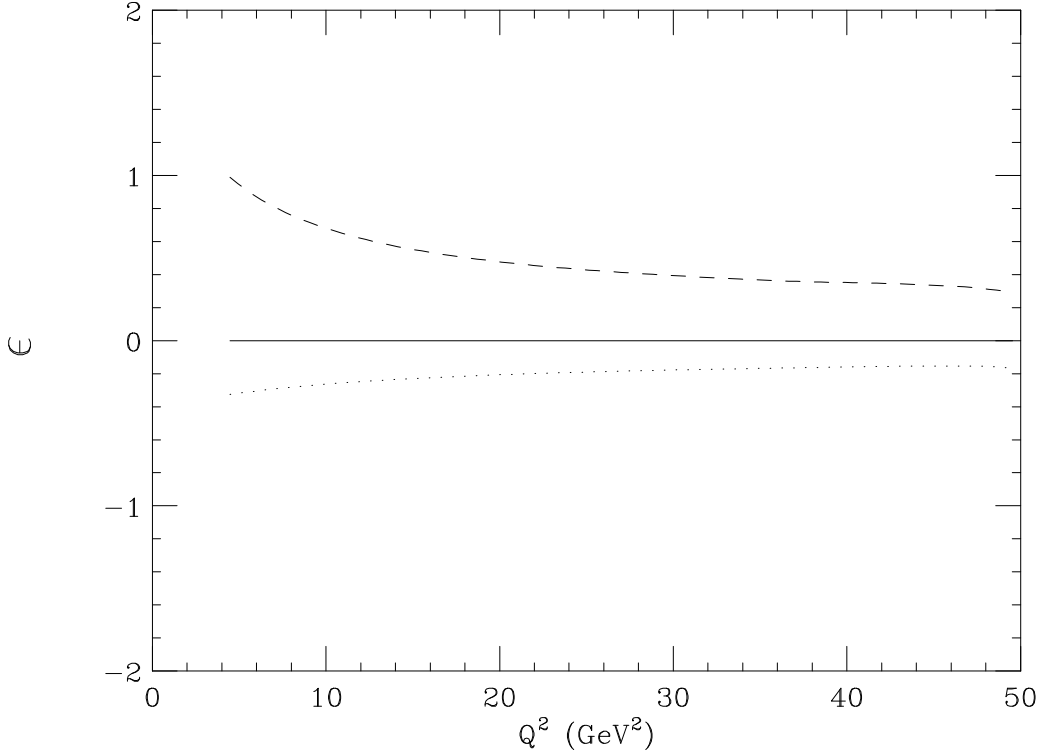}\hfill }
\caption{Theoretical error due to uncertainty in charm quark mass as a function of $Q^{2}$.  Error is plotted for $m_{c}=1.3 \; \mbox{GeV}$ and  $m_{c}=1.7 \; \mbox{GeV}$. }
\label{masserror}
\end{figure}

In Fig.~\ref{loQmc}, we study the error in our prediction for the $O(\alpha_s)$ color-octet differential cross section resulting from uncertainty in the determination of the charm quark mass. The three histograms in the figure correspond to three different choices of $m_{c}$: $m_{c}=1.3 \;
\mbox{GeV}$ (dotted histogram), $m_{c}=1.5 \; \mbox{GeV}$ (solid
histogram), and  $m_{c}=1.7 \; \mbox{GeV}$ (dashed histogram). The error in our prediction for the cross section, as defined by 
\[ \epsilon = \left( {d \sigma(m_c) \over d Q^2} - {d \sigma(m_c=1.5~{\rm GeV}) \over d Q^2}\right) /{d \sigma(m_c=1.5~{\rm GeV}) \over d Q^2}, \] 
is plotted in Fig.~\ref{masserror} for $m_c = 1.3~{\rm GeV}$ (dashed
line) and $1.7~{\rm GeV}$ (dotted line).  Though there is some
improvement as $Q^2$ is increased, the error is still $^{+60\%}_{-25
\%}$ at $Q^2 = 10~{\rm GeV}^2$. This is much greater than errors we
expect to receive from higher orders in perturbation theory and higher
twist effects, so the uncertainty in the charm quark mass dominates
the theoretical error in our prediction. 

\begin{figure}
\rotate[l]{
\epsfxsize=10cm
\hfil\epsfbox{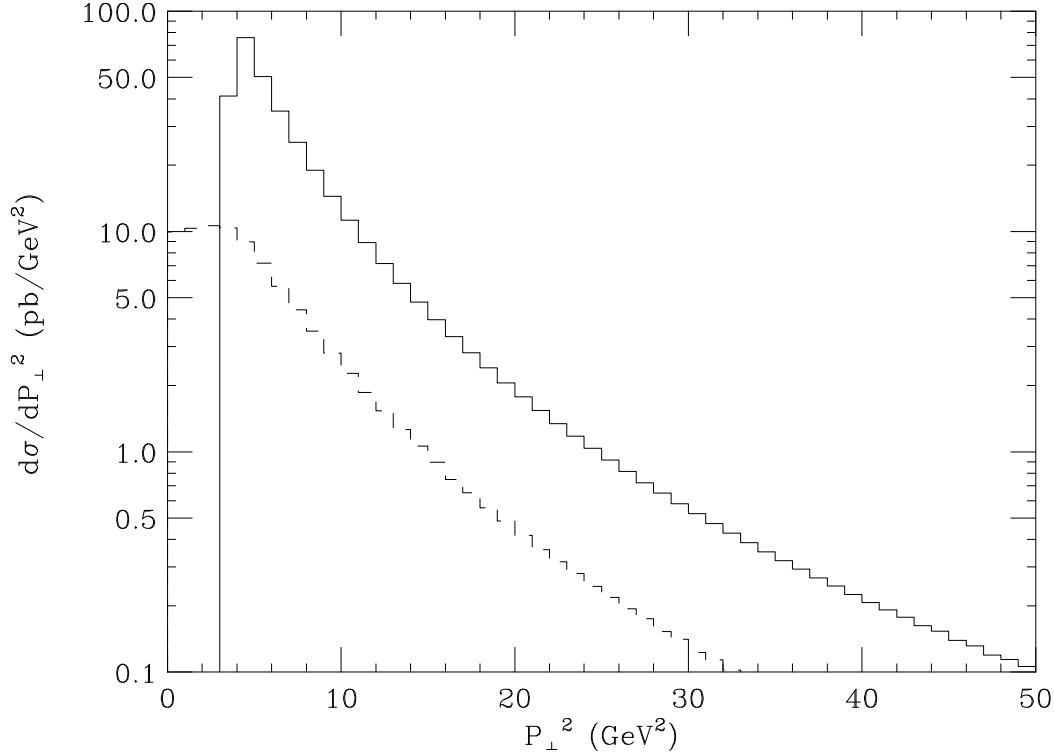}\hfill }
\caption{Differential cross section as a function of 
$P_{\perp}$. Solid - $O(\alpha_s)$ color-octet; dashed - $O(\alpha_s^2)$ color-singlet. We require  $Q^2 > 4 \; \mbox{GeV}^2$; no $P_{\perp}$ cut is imposed.}
\label{loPt}
\end{figure}

\begin{figure}
\rotate[l]{
\epsfxsize=10cm
\hfil\epsfbox{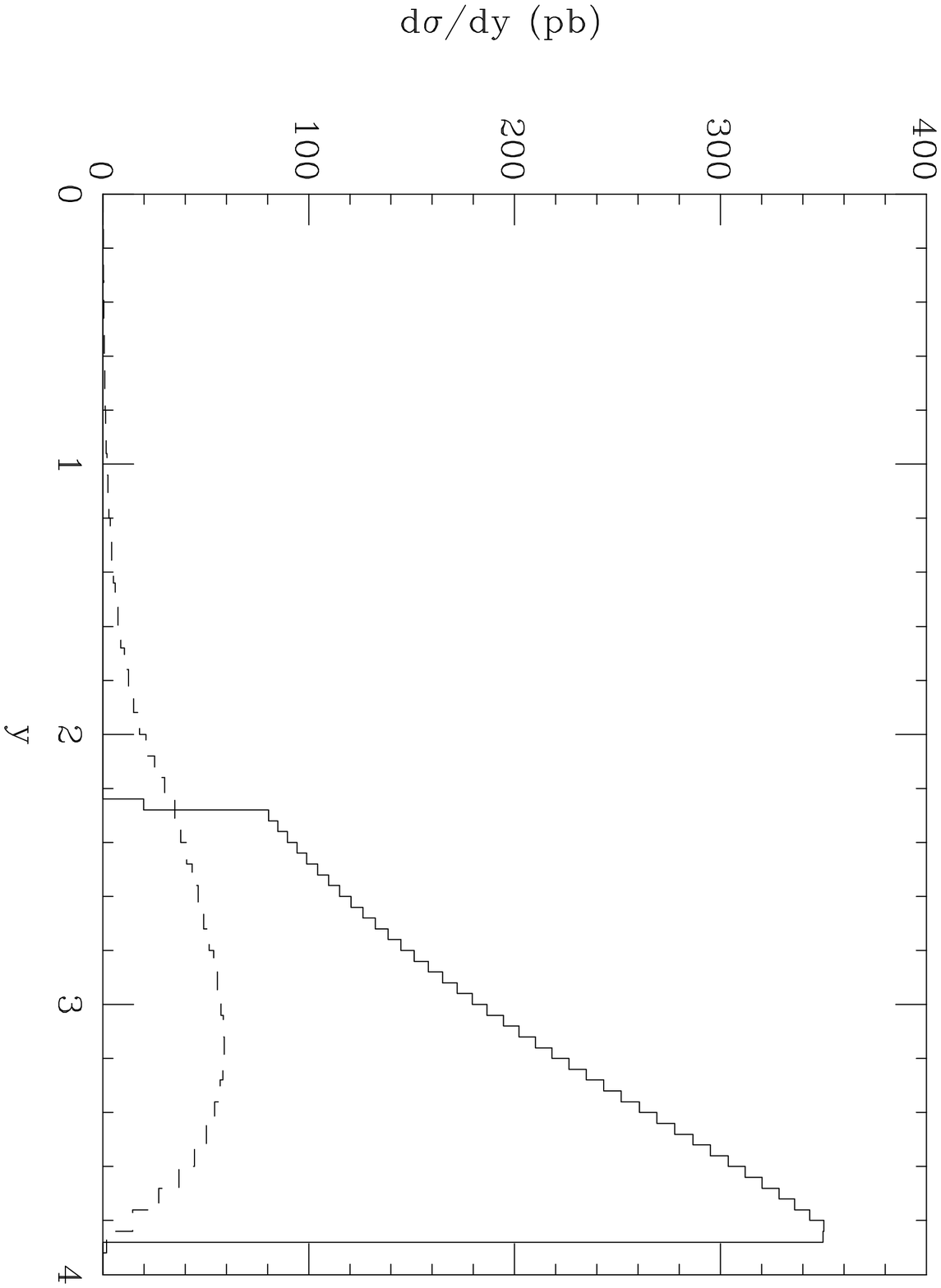}\hfill }
\caption{Differential cross section as a function of 
rapidity ($y$). Solid - $O(\alpha_s)$ color-octet; dashed - $O(\alpha_s^2)$ color-singlet. We impose the cut $Q^2 > 4 \; \mbox{GeV}^2$}
\label{lorap}
\end{figure}

In addition to the $Q^{2}$ distribution, we have also calculated the
$P^2_{\perp}$ distribution shown in Fig.~\ref{loPt}, and the rapidity
distribution shown Fig.~\ref{lorap}.  The rapidity of the $J/\psi$, is
defined to be: \[y = {1\over 2} {\rm ln} \left( {E+P_z\over E-P_z}
\right) ,\] where $E$ and $P_z$ are the energy and $z$-component of
the $J/\psi$ three-momentum measured in the photon-proton
center-of-momentum frame with the z-axis defined by the photon
three-momentum in that frame.  We only present results for one choice
of scale, $\mu^2 = Q^{2}+(2m_{c})^{2}$, one choice of charm quark
mass, $m_{c}=1.5 \; \mbox{GeV}$, and for $\langle {\cal
O}^{\psi}_8(^1S_0) \rangle = 0.01 \; \mbox{GeV}^3$, $\langle {\cal
O}^{\psi}_8(^3P_0)\rangle /m^2_c=0.005\;\mbox{GeV}^3$.  Note that in
Fig.~\ref{lorap} there is no $P_{\perp}$ cut. The sharp fall off of
the distribution at $P^2_{\perp} \sim 3 \; \mbox{GeV}^2$ is due to the
$Q^2 > 4 \; \mbox{GeV}^2$ cut. The $P_{\perp}^2$ and $y$ distributions
suffer considerable error due to the charm quark mass uncertainty.  In
the case of the $P_{\perp}^2$ distribution, the errors are similar in
size to the $Q^2$ distribution, and, like the $Q^2$ distribution,
decrease slightly as $P_{\perp}^2$ is increased. For the $y$
distribution, the error is roughly $^{+50\%}_{-25\%}$, and is
independent of $y$.

\begin{figure}
\rotate[l]{
\epsfxsize=10cm
\hfil\epsfbox{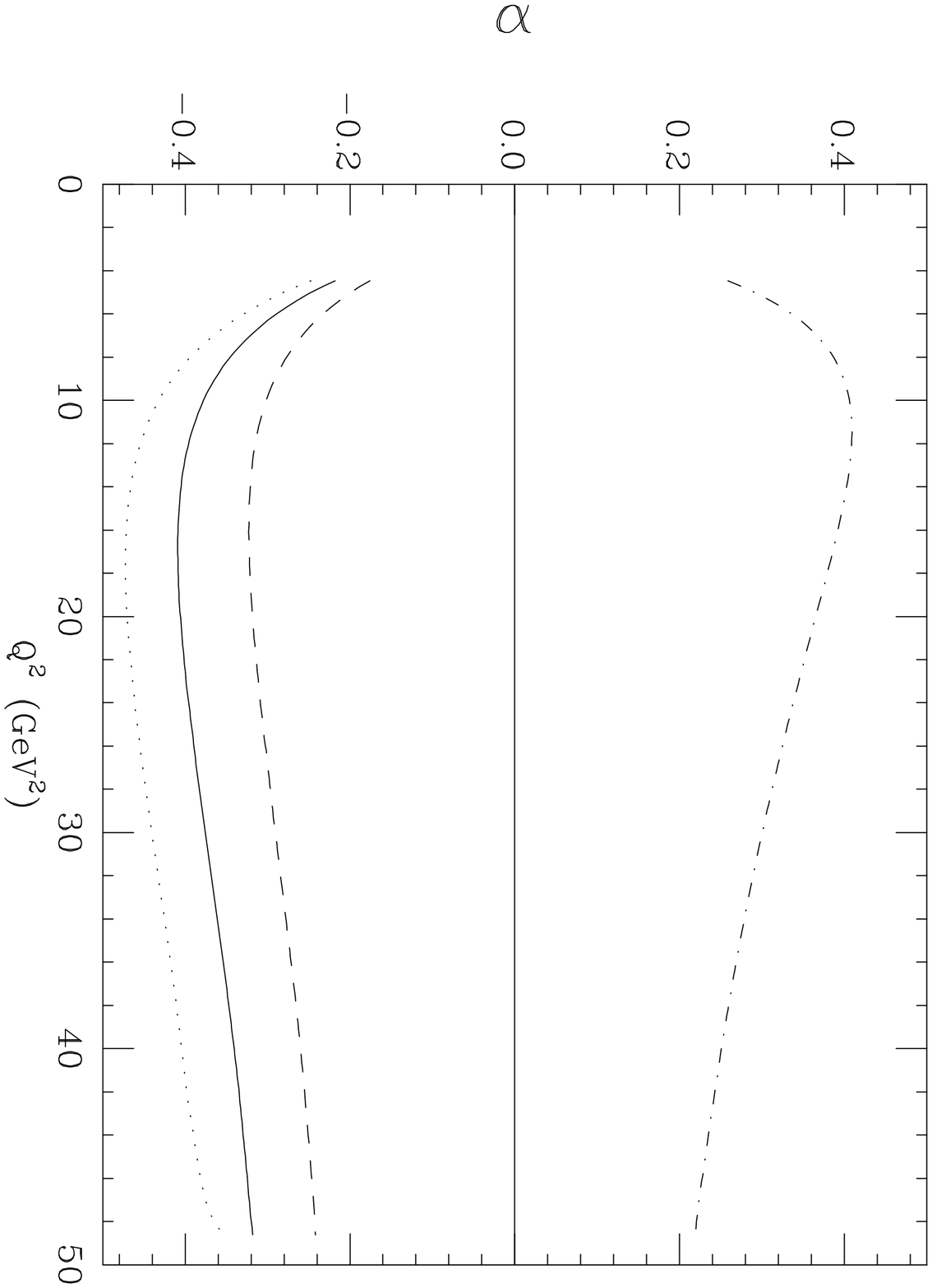}\hfill }
\caption{Polarization parameter $\alpha$ as function of $Q^2$. Dashed 
line - R = 0.25; Solid line - R = 0.5; Dotted line - R = 1.0;
Dashed-dotted line - R = -0.075}
\label{Alpha}
\end{figure}

\begin{figure}
\rotate[l]{
\epsfxsize=10cm
\hfil\epsfbox{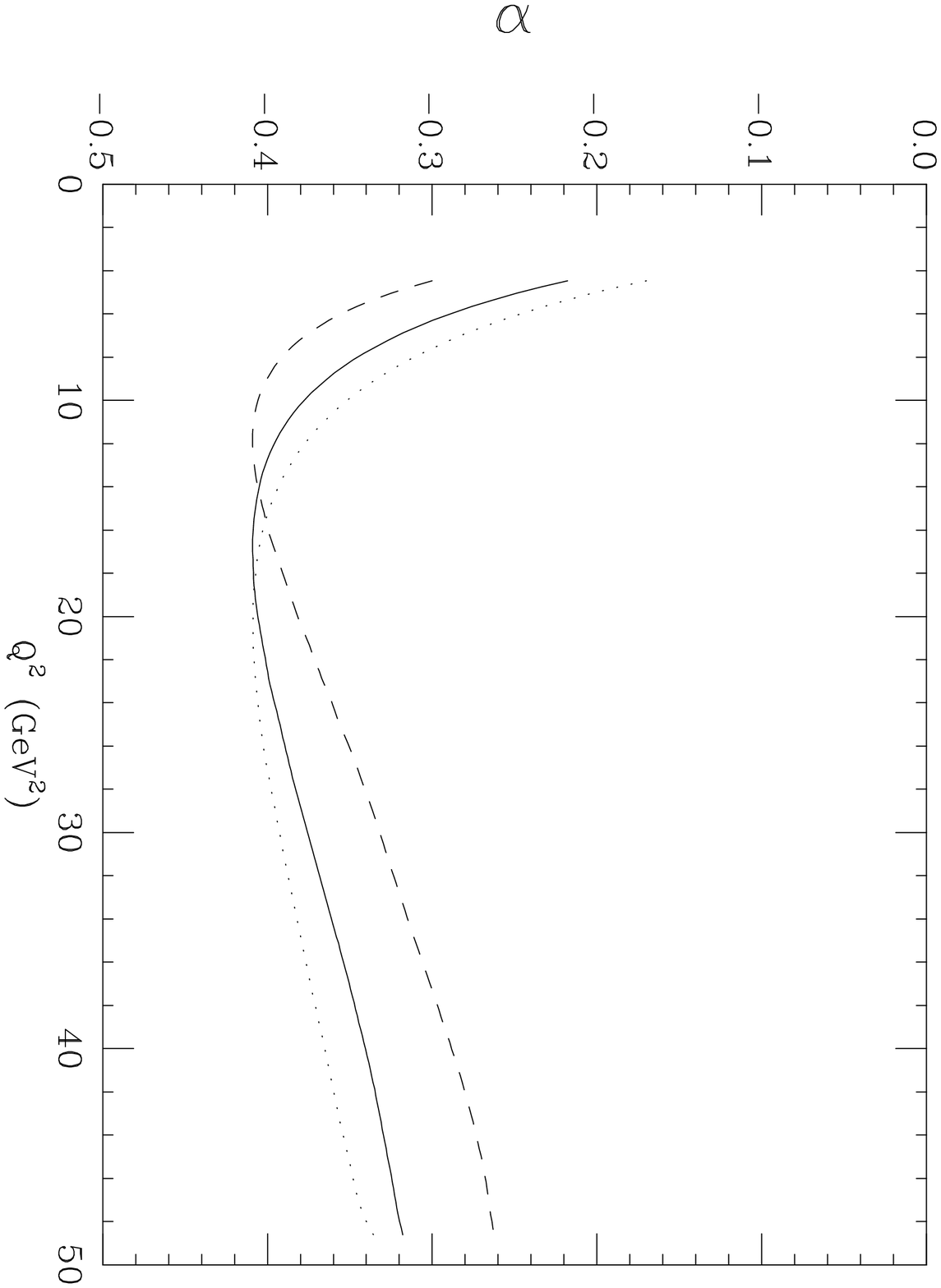}\hfill }
\caption{Polarization parameter $\alpha$ as function of $Q^2$. R = 0.5;  Solid line - $m_c = 1.5 {\rm GeV}$; Dotted line - $m_c = 1.7 {\rm GeV}$; Dashed line - $m_c = 1.3 {\rm GeV}$.}
\label{mcpol}
\end{figure}

Once the color-octet matrix elements have been determined from
leptoproduction of unpolarized $J/\psi$, it is possible to make
definite predictions for the production of polarized $J/\psi$. The
expression for the polarization parameter, $\alpha$, is given in
Eq.~(\ref{polparm}) and depends only on the ratio of color-octet
matrix elements $R \equiv \langle {\cal O}^{\psi}_8(^3P_0)
\rangle /(m_c^2 \langle {\cal O}^{\psi}_8(^1S_0) \rangle)$.
In Fig.~\ref{Alpha}, we plot $\alpha$ as a function of $Q^2$,
choosing $\mu^2 = Q^{2}+(2m_{c})^{2}$, and  $m_{c}=1.5 \; \mbox{GeV}$. 
Plots are shown for $R = 1.0, 0.5, 0.25, -0.075$. If the $c\bar{c}$ 
produced in the short distance process is in a $^1S_0$ state, the $J/\psi$ will be unpolarized, so we expect $\alpha = 0$ when $R =0$. Positive values of R lead to slightly longitudinally polarized $J/\psi$ ($\alpha < 0$), while if 
$R < 0$ (i.e. $\langle {\cal O}^{\psi}_8(^3P_0) \rangle < 0$)
the $J/\psi$ emerges with slightly transverse polarization. The error due to the uncertainty in the charm quark mass is significant, but not as large as the errors in the $P_{\perp}$ or $y$ distributions. To illustrate this, we plot the polarization in the case of $R = 0.5$ for three different values of $m_c$ in Fig.~\ref{mcpol}. The error is largest at large values of $Q^2$ where it is approximately $^{+20\%}_{-5\%}$.  Note that we have only included the $O(\alpha_s)$ color-octet contribution.  Since the $O(\alpha_s^2)$ color-singlet mechanism contributes about 25\% of the total cross section, its effect on the polarization needs to be included before detailed comparison with experiment can be made.  


In summary, the total cross section for
leptoproduction of $J/\psi$ is dominated by leading order color-octet
terms; the $O(\alpha_s^2)$ color-singlet term contributes only 
20-25\% of the total cross section. An accurate
measurement of the distribution $d\sigma/dQ^2$ will, thus, allow for an
independent determination of $\langle {\cal O}_8^{\psi}(^1S_0)\rangle$ and 
$\langle {\cal O}_8^{\psi}(^3 P_0)\rangle$. 
As discussed in section IV, nonperturbative
errors are expected to be small at large $Q^2$. For $Q^2 > 4 \;
\mbox{GeV}^2$ higher twist effects result in an error of less than 
8\%. In addition we
estimate the error due to the scale uncertainty; we find it is less
than 10\%. Finally we estimate the error due to the uncertainty in the
determination of the charm quark mass, and find it to be the dominant
error. One must require $Q^2 > 10 \;\mbox{GeV}^2$ for the error to be
less than 50\%. Errors in the extraction of
$\langle {\cal O}_8^{\psi}(^1S_0)\rangle$ and $\langle
{\cal O}_8^{\psi}(^3P_0)\rangle$ from photoproduction ($Q^2 = 0$) are
obviously greater than $100\%$. The errors that arise in extractions
from hadroproduction reported in Ref. \cite{BK} are greater than
100\%, and do not even include error due to uncertainty in the charm
quark mass. Therefore, measurement of leptoproduction of $J/\psi$ at
$Q^2 \geq 10~{\rm GeV^2}$ will be a great improvement over existing
extractions. Nevertheless, it is somewhat disappointing that our
prediction is dominated by the error due to the uncertainty in the
charm quark mass since it cannot be
systematically reduced by doing higher order perturbative QCD
calculations.

Once $\langle {\cal O}_8^{\psi}(^1S_0)\rangle$ and 
$\langle {\cal O}_8^{\psi}(^3P_0)\rangle$ are
determined, distributions in $P^2_{\perp}$ and rapidity, as well as the
polarization of the $J/\psi$, can also be predicted. The polarization
of $J/\psi$ is least sensitive to errors associated with the uncertainty
in the charm quark mass. Thus, an experimental measurement of the
polarization will be an important test of the NRQCD factorization
formalism.

In addition, we have shown that the production of $J/\psi$ with a well
separated gluon jet occurs 
only at $O(\alpha_s^2)$, and makes a small contribution to the total
cross section.  For these final states, the color-octet contribution
is suppressed, and only modifies the normalization,
but not the shapes of distributions. For this
reason, we do not feel that production of $J/\psi$ in association with
a hard jet will shed any light on color-octet mechanisms.

\section{Conclusion}

In this paper, we studied leptoproduction of $J/\psi$ at large
$Q^2$, and found that the cross section is dominated by
$O(\alpha_s)$ color-octet production mechanisms. Thus, by measuring the
total cross section as a function of $Q^2$, it is possible to
obtain measurements of the color-octet matrix elements 
$\langle {\cal O}_8^{\psi}(^1S_0)\rangle$ and 
$\langle {\cal O}_8^{\psi}(^3P_0)\rangle$. Neither of these has been
determined to better than an order of magnitude. 
The dominant error is due to the uncertainty in the value of the charm
quark mass. This results in roughly a 50\% uncertainty in the
prediction for the differential cross section for $Q^2 > 10 \; \mbox{GeV}^2$. 
We, therefore, estimate that $\langle {\cal O}_8^{\psi}(^1S_0)\rangle$
and $\langle {\cal O}_8^{\psi}(^3P_0)\rangle$ can be measured in
$J/\psi$ leptoproduction at about the 50\% level. 

In addition, we pointed out that once the color-octet matrix elements
have been measured it is possible to make a parameter free prediction
of the polarization of leptoproduced $J/\psi$. This will provide an
important test of the NRQCD factorization formalism. 

As a result of the work presented in this paper we conclude that
leptoproduction of unpolarized $J/\psi$ at $Q^2 > 10 \; \mbox{GeV}^2$ 
will provide a much better determination of the color-octet matrix
elements $\langle {\cal O}_8^{\psi}(^1S_0)\rangle$ and 
$\langle {\cal O}_8^{\psi}(^3P_0)\rangle$ than currently available. In
addition a measurement of the polarization of leptoproduced $J/\psi$
will provide an important test of the NRQCD factorization
formalism. We would, therefore, encourage an experimental measurement
of $J/\psi$ leptoproduction at large $Q^2$.

\bigskip

We would like to thank Adam Falk for countless consultations,
suggestions, and for reading over the final drafts of the paper. In
addition we would like to thank David Rainwater for help with 
the Monte Carlo integration routines.
The work of S.F. was supported in
part by the U.S.~Department of Energy under Grant
no.~DE-FG02-95ER40896, in part by the University of Wisconsin Research
Committee with funds granted by the Wisconsin Alumni Research
Foundation.
The work of T.M. was supported by the National Science Foundation
under Grant No. PHY-9404057. 

\pagebreak


\begin{references}

\bibitem{CDF} For a review of experimental aspects of quarkonia
production, see Sansoni, A. (CDF Collaboration), Nucl.\ Phys.\ {\bf 
A610}, 373c (1996).

\bibitem{Schuler} For an extensive review of the color-singlet model, 
see G. A. Schuler, CERN-TH-7170-94, hep-ph/9403387 (unpublished).

\bibitem{BBL} G.T. Bodwin, E. Braaten, and G.P. Lepage, Phys.\ Rev.\ 
D {\bf 51} 1125 (1995).

\bibitem{lmnmh} G.P. Lepage, L. Magnea, C. Nakhleh, U. Magnea, and K. 
Hornbostle, Phys.\ Rev.\ D {\bf 46}, 4052 (1992).

\bibitem{CL} P. Cho and A. Leibovich, Phys.\ Rev.\ D {\bf 53}, 150
(1996); {\bf 53}, 6203 (1996).

\bibitem{BK} M. Beneke and M. Kr$\ddot{\rm{a}}$mer, Phys.\ Rev.\ D
{\bf 55}, 5269 (1997); M. Beneke, Report No. 
CERN-TH/97-55, hep-ph/9703429 (unpublished)

\bibitem{Photo} J. Amundson, S. Fleming, and I. Maksymyk, Report No. 
UTTG-10-95, hep-ph/9601298 (unpublished); M. Cacciari and M.
Kr$\ddot{\rm{a}}$mer, Phys.\ Rev.\ Lett.\ {\bf 76}, 4128 (1996); P. 
Ko, J. Lee, and H.S. Song, Phys.\ Rev.\ D {\bf 54}, 4312 (1996).

\bibitem{Hadro} M. Beneke and I. Rothstein, Phys.\ Rev.\ D {\bf 54}, 
2005 (1996); S. Gupta and R. Sridhar, {\it ibid.} {\bf 54}, 5455
(1996); {\it ibid.} {\bf 55}, 2650 1997; W.K. Tang and M.
V$\ddot{\rm{a}}$nttinen, {\it ibid.} {\bf 54}, 4349 (1996); {\it
ibid.} {\bf 53}, 4851 (1996).

\bibitem{IR}  H.W. Huang and K.T. Chao, Phys.\ Rev.\ D {\bf 55} 244, 
(1997); E. Braaten and Y.Q. Chen, Phys.\ Rev.\ D {\bf 55}, 7152 (1997).

\bibitem{BC} E. Braaten and Y. Chen, Phys.\ Rev. {\bf D54}, 3216
(1996);
S. Fleming and I. Maksymyk, {\it ibid.} {\bf 54}, 3608 (1996).

\bibitem{mmm} H. Merabet, J.-F. Mathiot, R. Mendez-Galain, Z.\ Phys C 
62, 639 (1994).

\bibitem{HERA} Cartiglia, Nicolo, Report No. hep-ph/97032245
(unpublished).

\bibitem{Diffractive} S. J. Brodsky, L. Frankfurt, J. F. Gunion, A.H. 
Mueller, and M. Strikman, Phys.\ Rev.\ D {\bf 50}, 3134 (1994);
M.G. Ryskin, R.G. Roberts, A.D. Martin, and E.M. Levin, Report No. RAL-TR-95-065, hep-ph/9511228 (unpublished).

\bibitem{H1} H1 Collaboration, S. Aid, {\it et. al.}, Nucl. Phys.{\bf 
B468}, 3 (1996).

\bibitem{Factor} J.~C.~Collins, D.~E.~Soper, Ann.\ Rev.\ Part. Sci.\ 
37, (1987) 383.

\bibitem{Twist} J.~P.~Ma, Report No.~RCHEP-96/07, hep-ph/9705445,
(unpublished).

\bibitem{Shape} M.~Beneke, I.~Z.~Rothstein, and M.~B.~Wise, Report
No.~CERN-TH/97-86, hep-ph/9705286, (unpublished).

\bibitem{bsk} G.T. Bodwin, D.K. Sinclair, and S. Kim, Phys.\ Rev.\
Lett.\ {\bf77}, 2376 (1996).

\bibitem{EMC} EMC Collaboration, J.J. Aubert, {\it et. al.}, Nucl.
Phys. {\bf B213}, 1 (1983); D. Allasia {\it et al}. Phys. Lett. {\bf 
258B}, 493 (1991); Ch. Moriotti, Nucl. Phys. {\bf A532}, 437 (1991).

\bibitem{SF} S. Fleming, Report No. hep-ph/9610372 (unpublished).

\end{references}
\end{document}